\documentclass{article}
\usepackage{graphicx} 
\usepackage[utf8]{inputenc}
\usepackage[]{amsthm} 
\usepackage{caption}
\usepackage[]{amssymb} 
\usepackage[]{amsmath,txfonts}
\usepackage{float}
\usepackage{longtable}
\usepackage{siunitx} 
\usepackage{bm}
\usepackage{multirow}
\usepackage{graphicx}
\graphicspath{ {./images/} }
\usepackage[toc,page]{appendix}
\usepackage{lscape}
\usepackage{multirow}
\usepackage{booktabs}
\usepackage{array}    
\usepackage[linesnumbered,ruled,vlined]{algorithm2e}
\usepackage{hyperref}
\hypersetup{
    colorlinks=true, allcolors = blue
    }
\usepackage{authblk}
\usepackage{stackengine}
\usepackage[style=apa]{biblatex}
\title{Beyond Arbitrary Replications: A Principled Approach to Simulation Design in Causal Inference}
\author[1]{Hugo Gobato Souto}
\affil[1]{\stackunder{{\stackunder{Institute of Mathematics and Computer Sciences at University of São Paulo, Brazil}{Av. Trab. São Carlense 400, 13566-590 São Carlos (SP), Brazil}}}{\stackunder{{hgsouto@usp.br}. {https://orcid.org/0000-0002-7039-0572}}}}
\author[2]{Francisco Louzada Neto}
\affil[2]{\stackunder{{\stackunder{Institute of Mathematics and Computer Sciences at University of São Paulo, Brazil}{Av. Trab. São Carlense, 400, São Carlos, 13566-590, Brazil}}}{\stackunder{{louzada@icmc.usp.br}. {https://orcid.org/0000-0001-7815-9554}}}}
\date{}

\begin{document}

\maketitle

\begin{abstract}
Evaluation of novel treatment effect estimators frequently relies on simulation studies lacking formal statistical comparisons and using arbitrary numbers of replications ($J$). This hinders reproducibility and efficiency. We propose the Test-Informed Simulation Count Algorithm (TISCA) to address these shortcomings. TISCA integrates Welch's t-tests with power analysis, iteratively running simulations until a pre-specified power (e.g., 0.8) is achieved for detecting a user-defined minimum detectable effect size (MDE) at a given significance level ($\alpha$). This yields a statistically justified simulation count ($J$) and rigorous model comparisons. Our bibliometric study confirms the heterogeneity of current practices regarding $J$. A case study revisiting \citeauthor{McJames2024} (\citeyear{McJames2024}) demonstrates TISCA identifies sufficient simulations ($J=500$ vs. original $J=1000$), saving computational resources while providing statistically sound evidence. TISCA promotes rigorous, efficient, and sustainable simulation practices in causal inference and beyond.
\end{abstract}

\section{Introduction}

In the field of causal inference, estimating the effect of a treatment or intervention is of paramount importance across a variety of disciplines, including economics \parencite{https://doi.org/10.48550/arxiv.1812.04345,Varian2016,Abadie2010,Card1999}, epidemiology \parencite{Rothman2005,Ohlsson2020,Vandenbroucke2016}, and social sciences \parencite{Yeager2019,Yeager2022,Bail2019}. Two fundamental quantities of interest in this context are the Average Treatment Effect (ATE) and the Conditional Average Treatment Effect (CATE). These metrics provide critical insights into the impact of a treatment on a population, both at an aggregate level and within specific subgroups defined by covariates. 

Common approaches to estimating both CATE and ATE include tree-based methods such as Causal Regression Forests \parencite{Wager2018}, Bayesian Additive Regression Trees (BART) \parencite{Chipman2010}, Bayesian Causal Forest (BCF) \parencite{Hahn2020} and other machine learning techniques, such as the neural network models TNet \parencite{pmlr-v130-curth21a}, TARNet \parencite{pmlr-v70-shalit17a}, DragonNet \parencite{NEURIPS2019_8fb5f8be}, DR-CFR \parencite{Hassanpour2020Learning}, and SNet \parencite{pmlr-v130-curth21a}. These nonparametric models can capture complex interactions between covariates and even estimate the treatment assignment mechanism/distribution. Almost every month, novel and state-of-the-art models are proposed in the literature, advancing the field of treatment effect estimation. Nonetheless, the current practices in models evaluation and comparisons are far from perfect and could make use of more statistical rigours. 

Due to the inherent challenges in causal inference, particularly the unobservability of counterfactual outcomes, researchers resort to using synthetic and semi-synthetic data for model evaluation \parencite{BENCHMARKS2021_2a79ea27}. These data allow for controlled experiments where the true treatment effects are known, thus enabling the assessment of model performance in estimating the ATE and CATE \parencite{BENCHMARKS2021_2a79ea27}. Synthetic data are fully simulated datasets where both the covariates $\mathbf{X}$ and the potential outcomes $Y(1)$ and $Y(0)$ are generated according to pre-specified distributions and relationships. The advantage of using synthetic data lies in the precise control it offers over the data generating process (DGP), allowing researchers to test models under various controlled scenarios \parencite{BENCHMARKS2021_2a79ea27}. Formally, let the DGP be defined as follows:

\begin{equation}
Y_i(Z_i) = f(\mathbf{X}_i) + \tau(\mathbf{X}_i) \cdot Z_i + \epsilon_i
\end{equation}

where $f(\mathbf{X}_i)$ is the baseline outcome model, $\tau(\mathbf{X}_i)$ is the true CATE, and $\epsilon_i$ is the error term. By specifying $f(\mathbf{X})$, $\tau(\mathbf{X})$, and the distribution of $\epsilon$, researchers can generate the potential outcomes $Y(1)$ and $Y(0)$ for any given covariate vector $\mathbf{X}$.

Semi-synthetic data, on the other hand, combine real-world covariates $\mathbf{X}$ with synthetic potential outcomes generated under a specified DGP. This approach retains the realistic covariate distribution of actual datasets while allowing for controlled experimentation with the outcome-generating process \parencite{BENCHMARKS2021_2a79ea27}. The use of semi-synthetic data is particularly appealing when real-world covariates exhibit complex dependencies that are difficult to capture through fully synthetic simulations.

Nevertheless, it is worth mentioning that both the use of synthetic and semi-synthetic data when evaluation novel models can be problematic if not done meticulously and considering the DGPs biases and limitations \parencite{BENCHMARKS2021_2a79ea27}. In their paper, \citeauthor{BENCHMARKS2021_2a79ea27} (\citeyear{BENCHMARKS2021_2a79ea27}) address these problems found in many studies in the literature of models for treatment effect estimation and propose concrete actions that researchers ought to undertake to prevent such issues. Similar to the paper of  \citeauthor{BENCHMARKS2021_2a79ea27} (\citeyear{BENCHMARKS2021_2a79ea27}), this paper puts light on problems found in the current analysis practices of model evaluation for treatment effect estimation and proposes solutions to avoid such issues. But before stating these problems, we must first understand the current analysis practices of model evaluation for treatment effect estimation.

The evaluation of model performance in the context of CATE and ATE estimation typically involves metrics such as the Root Mean Squared Error (RMSE), Coverage, and Confidence Interval Length (CIL) \parencite{Hahn2020,Hill2011}. For frequentist models, RMSE is the predominant metric used to assess the accuracy of the estimated treatment effects \parencite{pmlr-v130-curth21a,Hill2011}. RMSE for CATE is defined as:

\begin{equation}
\text{RMSE}_{CATE} = \sqrt{\frac{1}{n} \sum_{i=1}^{n} (\hat{\tau}_i - \tau_i)^2}
\end{equation}

where $\hat{\tau}_i$ is the estimated treatment effect for unit $i$, and $\tau_i$ is the true treatment effect. It is worth mentioning that $\text{RMSE}_{CATE}$ is more commonly referred to as Precision in Estimation of Heterogeneous Effect (PEHE) in the literature. The RMSE for ATE, on the other hand, is given as:

\begin{equation}
\text{RMSE}_{ATE} = \sqrt{\frac{1}{J} \sum_{i=j}^{J} (\hat{\tau}_j - \tau_j)^2}
\end{equation}

where $\hat{\tau}_j$ is the estimated average treatment effect for simulation/dataset $j$, and $\tau_j$ is the true average treatment effect.

For Uncertainty quantification models, especially Bayesian models, additional metrics such as Coverage and CIL are commonly used \parencite{Hahn2020,Hill2011}. For CATE, coverage refers to the proportion of times the true treatment effect $\tau_i$ falls within the estimated $100(1-\alpha)\%$ credible interval:

\begin{equation}
\text{Coverage}_{CATE} = \frac{1}{n} \sum_{i=1}^{n} \mathbf{1} \left( \tau_i \in \left[ \hat{\tau}_i^{\text{lower}}, \hat{\tau}_i^{\text{upper}} \right] \right)
\end{equation}

where $\mathbf{1}(\cdot)$ is the indicator function, and $\left[ \hat{\tau}_i^{\text{lower}}, \hat{\tau}_i^{\text{upper}} \right]$ is the credible interval for $\tau_i$. For ATE,  coverage refers to the proportion of times the true average treatment effect $\tau_j$ for simulation/dataset $j$ falls within the estimated $100(1-\alpha)\%$ credible interval:
\begin{equation}
\text{Coverage}_{ATE} = \frac{1}{J} \sum_{i=j}^{J} \mathbf{1} \left( \tau_j \in \left[ \hat{\tau}_j^{\text{lower}}, \hat{\tau}_j^{\text{upper}} \right] \right)
\end{equation}

CIL, on the other hand, measures the width of these credible intervals for ATE and for CATE, the average width:

\begin{equation}
\text{CIL}_{CATE} = \frac{1}{n} \sum_{i=1}^{n} \left( \hat{\tau}_i^{\text{upper}} - \hat{\tau}_i^{\text{lower}} \right)
\end{equation}

\begin{equation}
\text{CIL}_{ATE}  = \left( \hat{\tau}_j^{\text{upper}} - \hat{\tau}_j^{\text{lower}} \right)
\end{equation}

A shorter CIL indicates a more precise estimate, but this must be balanced against coverage to ensure that the intervals are not too narrow.

While these metrics provide quantitative assessments of performance on a \textit{single} simulation run or dataset, the critical step involves aggregating results across multiple Monte Carlo simulations (indexed by $j=1, \dots, J$ in the definitions above) to compare the overall performance of different models, particularly when evaluating a newly proposed estimator against existing benchmarks. It is precisely within this aggregation and comparison process that two significant methodological issues arise in the current literature on treatment effect estimation models.

First, there is a prevalent lack of formal statistical testing when comparing the performance of a novel proposed model against established benchmarks using these aggregated metrics. Often, conclusions about model superiority are drawn based on observing seemingly lower average RMSE or better average coverage across the $J$ simulations, without statistically assessing whether these observed differences are likely due to genuine performance advantages or simply random variation inherent in the Monte Carlo process. For instance, a new model might show a slightly lower average PEHE than a benchmark across $J$ simulations, but without a statistical test (e.g., a t-test comparing the distributions of PEHE values obtained from the two models across the simulations), it remains uncertain whether this difference is statistically significant or could have occurred by chance \parencite{Souto2024,MCS,Hollander2015,Witt2003}.

Second, and closely related to the first issue, is the often arbitrary selection of the number of simulation replications, $J$. As our bibliometric analysis in Section 2 will demonstrate, the choice of $J$ varies considerably across studies, frequently ranging from as low as 5 simulations to several thousand, often based on convention, available computational resources, or an 'educated guess' rather than a principled, statistically motivated approach. This heterogeneity makes cross-study comparisons challenging and raises concerns about the reliability and efficiency of the findings \parencite{Souto2024,MCS,Hollander2015,Witt2003}. An insufficient number of simulations ($J$) may lead to underpowered comparisons, failing to detect genuine, meaningful differences between models (a Type II error), even if a statistical test were applied \parencite{Hollander2015,duPrel2010,Greenland2016,Hedges2001}. Conversely, performing an unnecessarily large number of simulations wastes valuable computational resources and time, hindering research productivity and increasing the environmental footprint of computational research.

These methodological shortcomings are particularly concerning in the current era, characterized by a rapid acceleration in research output, especially thanks to artificial intelligence (AI) and large language model (LLM) applications \parencite{Goldkuhle2024,doVale2024,https://doi.org/10.48550/arxiv.2408.15409,https://doi.org/10.48550/arxiv.2410.03019,https://doi.org/10.48550/arxiv.2204.08377}. The pressure to publish quickly, coupled with the ease of generating seemingly novel model variations, can lead to a proliferation of studies. Without rigorous evaluation standards, including formal statistical testing and justified simulation designs, we risk saturating the literature with papers claiming state-of-the-art performance based on statistically insignificant or underpowered findings \parencite{Goldkuhle2024,doVale2024,https://doi.org/10.48550/arxiv.2408.15409,https://doi.org/10.48550/arxiv.2410.03019,https://doi.org/10.48550/arxiv.2204.08377}. This not only hinders true scientific progress by creating noise and propagating potentially false conclusions, but it also erodes trust in the research community and misdirects efforts towards models that may not offer genuine improvements. Emphasizing methodological rigor, ensuring that claims of superiority are statistically sound, and promoting efficient use of computational resources are crucial steps to maintain research quality and integrity amidst this rapid expansion.

This paper addresses these two fundamental shortcomings in the evaluation methodology for treatment effect estimators. We propose a novel algorithm, named \textbf{T}est-\textbf{I}nformed \textbf{S}imulation \textbf{C}ount \textbf{A}lgorithm (TISCA), designed to integrate formal statistical hypothesis testing directly into the Monte Carlo simulation framework while simultaneously determining the necessary number of simulations required to achieve adequate statistical power. Our approach is grounded in the principles of power analysis, utilizing the Welch's t-test to compare model performance metrics (like RMSE) between a proposed model and its competitors.

Specifically, our algorithm requires the researcher to pre-specify three key parameters standard in hypothesis testing:
\begin{enumerate}
    \item The desired statistical power ($1 - \beta$), typically set at 0.80, representing the probability of detecting a true difference when it exists.
    \item The significance level ($\alpha$), usually set at 0.05, controlling the probability of a Type I error (falsely claiming a difference).
    \item The Minimum Detectable Effect size (MDE), representing the smallest difference in the chosen performance metric (e.g., difference in mean RMSE between the proposed model and the best benchmark) that the researcher deems practically meaningful and wishes to detect. This MDE can be informed by preliminary runs (e.g., 50 simulations) to gauge typical performance differences.
\end{enumerate}

Given these inputs, TISCA iteratively performs simulations, calculating the chosen performance metric for the models under comparison at each step. After each batch of simulations, it conducts a Welch's t-test and estimates the current statistical power based on the observed data and the specified MDE. The simulations continue until the estimated power reaches or exceeds the pre-specified target (e.g., 0.80).

Consequently, researchers employing our algorithm will not only obtain a statistically rigorous assessment of their model's relative performance against benchmarks but will also arrive at a justified and efficient number of simulation replications ($J$) tailored to their specific research question and desired level of certainty. This data-driven approach replaces arbitrary choices with a principled determination of simulation effort, thereby enhancing the credibility, reproducibility, and efficiency of model evaluation studies in treatment effect estimation.

The remainder of this paper is structured as follows: Section \ref{Sec2} presents a bibliometric study quantifying the heterogeneity in the number of simulations used in recent literature, empirically motivating the need for standardization. Section \ref{Sec3} details the proposed algorithm, discusses its theoretical underpinnings based on the Welch's t-test and power analysis, explores its strengths and limitations, and introduces a companion website with an interactive fine-tuned LLM tool designed to assist researchers in implementing the proposed algorithm (see \url{https://tisca-llm-app.streamlit.app/}). Section \ref{Sec4} provides a practical demonstration by revisiting the simulation study of \citeauthor{McJames2024} (\citeyear{McJames2024}), illustrating how our algorithm could have determined the necessary number of simulations, saving significant computational resources while ensuring statistical validity. Finally, Section \ref{Sec5} concludes with a summary of our contributions and discusses potential avenues for future research.

\section{Bibliometric Study} \label{Sec2}

\subsection{Study Design and Data Collection}
\label{subsec:biblio_methodology}

To empirically investigate the current practices regarding the number of simulation replications used in evaluating treatment effect estimation models, we conducted a targeted bibliometric analysis analysing a sample of 100 papers. The primary objective was to quantify the heterogeneity in the reported number of simulations ($J$) across relevant studies.

\subsubsection*{Search Strategy}
We utilized Google Scholar as the primary search database due to its broad coverage of academic literature, including peer-reviewed articles and preprints. The search was conducted using a combination of keywords relevant to the field of treatment effect estimation, specifically:
\begin{itemize}
    \item ``treatment effect estimation''
    \item ``heterogeneous treatment effect estimation''
    \item ``inverse probability weighting treatment effect estimation''
    \item ``nonparametric treatment effect estimation''
    \item ``matching treatment effect estimation''
\end{itemize}
The search focused on identifying papers published in recent years to capture contemporary practices (with 57\% of the papers being published in the last 5 years and 88\% in the last 10 years).

\subsubsection*{Selection Criteria}
From the initial pool of search results, studies were selected for inclusion based on the following criteria:
\begin{enumerate}
    \item The study must propose either a novel model for treatment effect estimation, a significant modification of an existing model aimed at improvement, or a novel methodological approach for estimating treatment effects (ATE or CATE).
    \item The study must include simulation experiments (using synthetic or semi-synthetic data) as part of its model evaluation or comparison methodology.
    \item The study must report the number of simulation replications ($J$) performed for these experiments.
\end{enumerate}
Studies that solely applied existing methods without proposing novelty, focused purely on theoretical aspects without simulations, or did not clearly report the number of simulations were excluded (with the later criterium culminating in only one exclusion, something positive from the considered literature).

\subsubsection*{Handling of Preprints}
Our sample included studies published in peer-reviewed journals as well as preprints hosted on arXiv. Preprints constitute a significant channel for disseminating cutting-edge research in machine learning and related fields, including treatment effect estimation. To ensure relevance and potential impact, arXiv preprints were included only if they met at least one of the following conditions:
\begin{itemize}
    \item The preprint had accrued at least one citation from a published paper (according to Google Scholar), suggesting some level of peer recognition or influence.
    \item The preprint had been posted within the last year prior to our data collection cutoff date, ensuring recency and allowing the inclusion of relevant papers that are presumably under review currently.
\end{itemize}
Approximately 10\% of our final sample consisted of such arXiv preprints. To assess the potential impact of including these non-peer-reviewed works, we performed a sensitivity analysis. We compared the distribution and summary statistics of the number of simulations ($J$) for the full sample with the results obtained when excluding the arXiv preprints. This comparison revealed that the overall findings and conclusions regarding the heterogeneity in simulation counts remained unchanged, confirming the robustness of our results irrespective of the inclusion of these preprints, which for some could be a controversial decision.

\subsubsection*{Final Sample}
Following the application of these search and selection criteria, a final sample of $N=100$ studies was compiled for analysis. These studies represent a cross-section of recent work in the field that utilizes simulation studies for evaluating novel contributions to treatment effect estimation. The distribution and characteristics of the reported number of simulations ($J$) across these studies are presented in the following subsection. 

\subsection{Results}
\label{subsec:biblio_results}

\begin{figure}[H]
\centering
\includegraphics[scale=0.55]{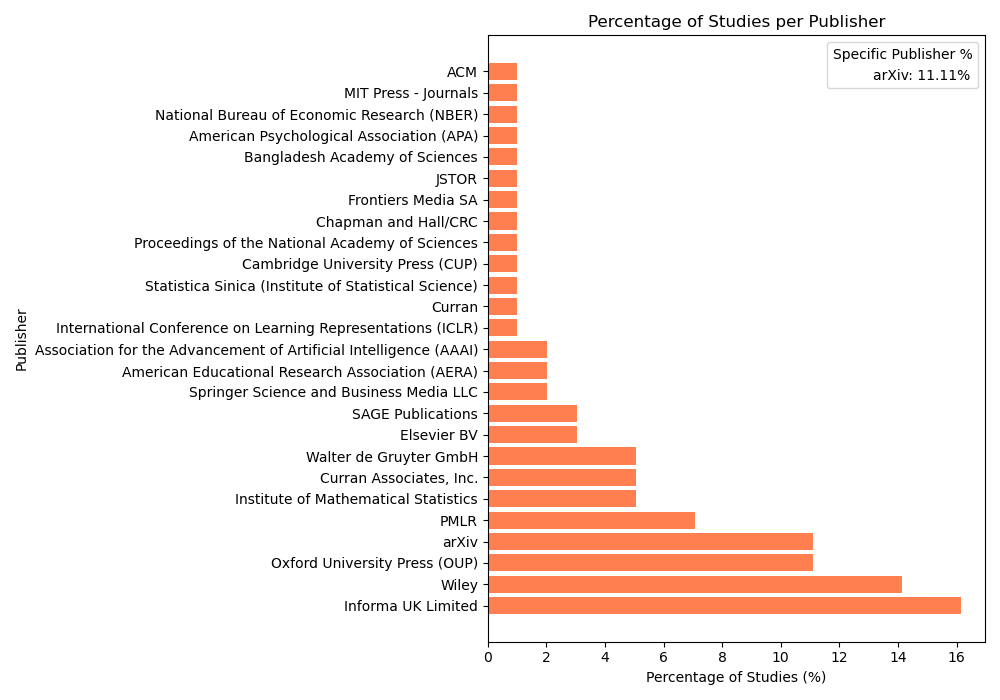}  
\caption{Percentage of Studies per Publisher}
\label{Figure1}
\centering
\end{figure}

\begin{figure}[H]
\centering
\includegraphics[scale=0.45]{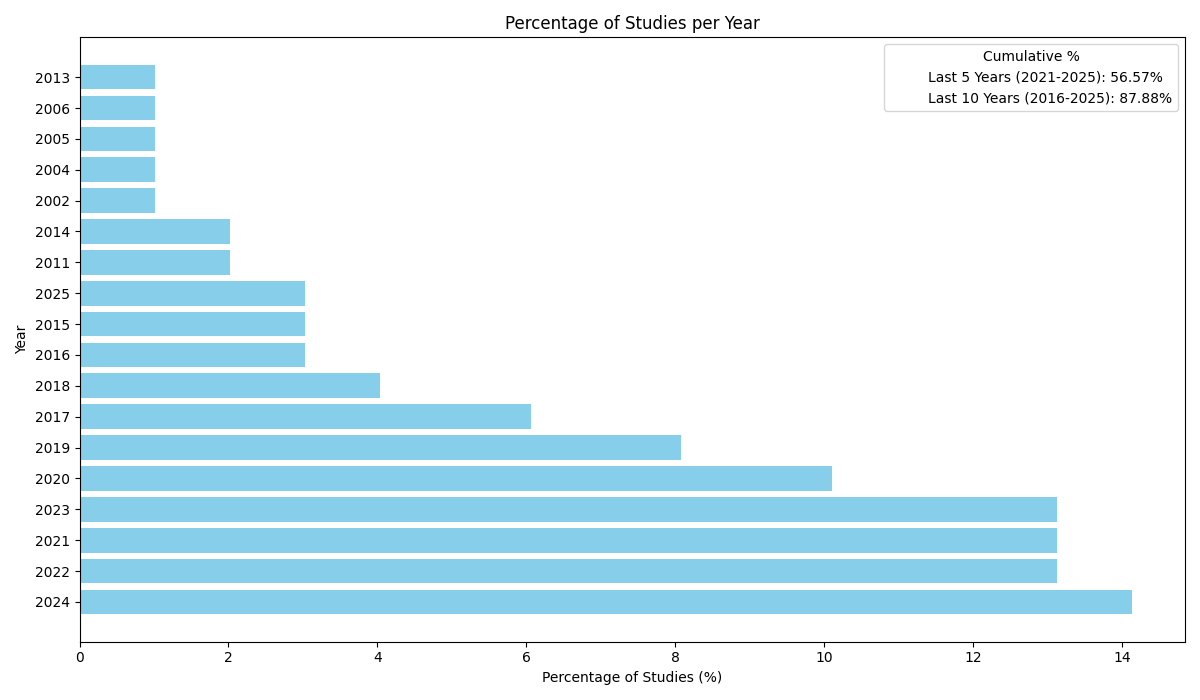}  
\caption{Percentage of Studies per Year}
\label{Figure2}
\centering
\end{figure}

\begin{figure}[H]
\centering
\includegraphics[scale=0.55]{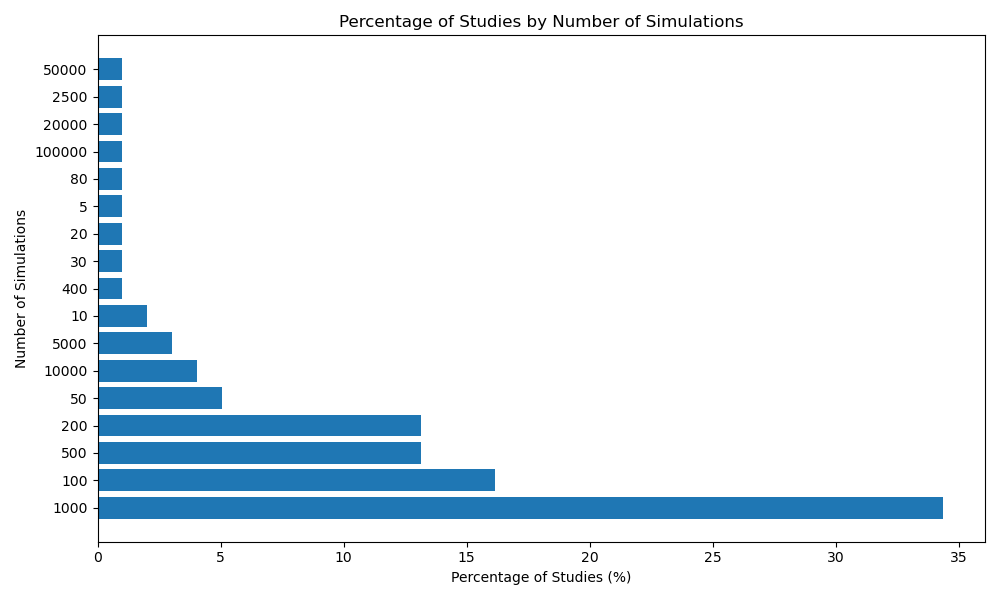}  
\includegraphics[scale=0.55]{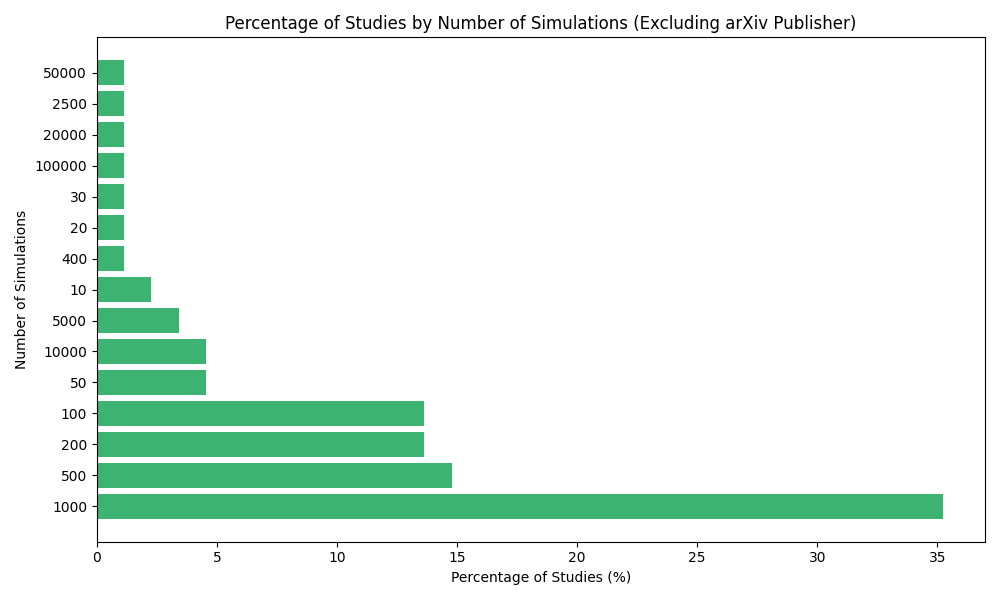} 
\caption{Percentage of Studies by Number of Simulations}
\label{Figure3}
\centering
\end{figure}

The results of the bibliometric analysis, conducted as described in Section~\ref{subsec:biblio_methodology}, are summarized in Figures~\ref{Figure1}, \ref{Figure2}, and \ref{Figure3}.

Figure~\ref{Figure1} illustrates the distribution of the 100 sampled studies across various publishers and publication outlets. The wide array of sources, ranging from top-tier statistics and machine learning journals (e.g., PMLR, Statistica Sinica) and conferences (e.g., ICLR, AAAI) to broader scientific venues (e.g., PNAS) and preprint servers (arXiv, representing 11.11\%), demonstrates that our sample is not confined to a narrow niche within the literature. 

Similarly, Figure~\ref{Figure2} shows the distribution of studies by publication year. While the sample spans over more than two decades, there is a clear concentration in recent years, with 56.57\% of studies published between 2021 and 2025 and 87.88\% published between 2016 and 2025. This temporal distribution ensures that our analysis reflects contemporary practices. Together, the diversity of publishers and the temporal spread underscore the representativeness and relevance of our sample, lending credibility to the findings regarding simulation practices.

The core finding of our bibliometric study is presented in Figure~\ref{Figure3}, which displays the distribution of the number of simulation replications ($J$) reported in the sampled studies. The top panel shows the distribution for the full sample ($N=100$), while the bottom panel shows the distribution after excluding arXiv preprints ($N=89$). Both plots reveal striking heterogeneity in the choice of $J$. The number of simulations ranges dramatically, from as few as 10 or even 5 replications in some studies to as many as 100,000 in others.

Despite this wide variation, a clear mode exists: $J=1000$ simulations is the most frequent choice, accounting for approximately 35\% of studies in both the full sample and the sample excluding arXiv publications. This confirms the robustness of this observation and highlights a common, albeit potentially unfounded, convention in the field. The pronounced heterogeneity, however, strongly indicates a lack of consensus or standardized, evidence-based guidelines for determining the appropriate simulation effort.

Crucially, simply adopting the most common practice ($J=1000$) as a de facto standard, which would be an alternative for the solution proposed in this paper, would be highly problematic and lack rigor. As indicated by the distributions in Figure~\ref{Figure3}, a substantial portion of the literature utilizes significantly fewer simulations. Specifically, 53.54\% of all studies (and 53.41\% of non-arXiv studies) in our sample used 500 or fewer simulations – meaning that adopting $J=1000$ as a standard would implicitly classify over half of the recent simulation studies analyzed as potentially insufficient, employing at least two times fewer replications than this arbitrary benchmark.

More fundamentally, selecting $J=1000$ merely based on its popularity is not rooted in any statistical or scientific principle related to the goals of the simulation study, such as achieving adequate statistical power for model comparison. It reflects convention rather than a rigorous assessment of need. This is precisely where the algorithm proposed in this paper offers a significant improvement. Instead of relying on arbitrary conventions or educated guesses, our approach provides a statistically principled method to determine the necessary number of simulations ($J$) based on researcher-specified parameters for significance ($\alpha$), power ($1-\beta$), and the minimum detectable effect size (MDE) for the comparison of interest. This ensures that the simulation effort is both justified and sufficient, promoting higher quality, more reliable, and more efficient research in the field of treatment effect estimation.

\section{The Proposed Algorithm: TISCA}
\label{Sec3}

Addressing the identified methodological gaps in the evaluation of treatment effect estimators---namely, the prevalent lack of formal statistical testing for model comparisons and the arbitrary selection of simulation replications---requires a systematic and statistically grounded approach. To this end, we introduce the \textbf{T}est-\textbf{I}nformed \textbf{S}imulation \textbf{C}ount \textbf{A}lgorithm (TISCA). TISCA is designed to integrate rigorous hypothesis testing directly into the Monte Carlo simulation workflow, simultaneously determining the minimum number of simulations ($J$) required to achieve a pre-specified statistical power for detecting meaningful differences in model performance.

\subsection{Methodological Foundation: The Welch's t-test}
\label{subsec:welch_test}

At the core of TISCA lies the comparison of performance metrics (e.g., PEHE, RMSE$_{ATE}$, Coverage, CIL) between the proposed model and the most performing benchmark model across $J$ simulation runs, which consequently means that if the proposed model is statistically significantly superior to the most performing benchmark model for a certain performance metric, it must also be superior to the other benchmark models. A natural choice for comparing the means of these metrics between two groups (e.g., performance of model A vs. model B across simulations) is the t-test \parencite{Welch1947}. However, a standard Student's t-test relies on the assumption of equal variances between the two groups being compared \parencite{Welch1947}.

During our bibliometric analysis, alongside general observations in simulation studies, we noticed that that the variability (standard deviation or variance) of performance metrics across Monte Carlo runs often differs between models, especially models that are highly different in nature (e.g., OLS linear model and Bayesian Causal Forest \parencite{Hahn2020}). For instance, a more complex model might exhibit higher variance in its PEHE estimates across simulations compared to a simpler, more stable benchmark. Violating the homogeneity of variance assumption can lead to unreliable p-values and potentially incorrect conclusions if a standard t-test is employed \parencite{Zimmerman2004}.

Therefore, TISCA utilizes the \textbf{Welch's t-test} \parencite{Welch1947} as its default comparison method. The Welch's t-test does not assume equal variances and adjusts its degrees of freedom accordingly using the Welch-Satterthwaite equation, providing a more robust comparison when the variances of the performance metrics between models are unequal.

A second assumption for the t-test (both Student's and Welch's) is the normality of the underlying data within each group. In our context, this refers to the distribution of a specific performance metric (e.g., PEHE values) obtained from the $J$ simulation runs for a given model. While true normality is seldom guaranteed, we leverage the \textbf{Central Limit Theorem (CLT)}. Many key performance metrics, such as RMSE$_{ATE}$ (an average error) and PEHE (average squared error), are derived from averaging or summing operations across units within each simulation. As the number of simulation replications ($J$) increases, the distribution of the mean of these metrics tends towards normality due to the CLT \parencite{dudley1978central,Brosamler1988}. We consider this assumption reasonably met, particularly for the sample sizes ($J$) typically encountered or targeted in simulation studies (e.g., $J >> 30$), which aligns with common heuristics for the applicability of the CLT \parencite{Kwiatkowski1992}.

Crucially, TISCA deliberately \textit{avoids} performing formal statistical tests for normality (e.g., Shapiro-Wilk test) prior to conducting the Welch's t-test. The practice of "pre-testing" assumptions like normality has been widely criticized in the statistical literature \parencite{Rochon2012, Zimmerman2004, Rasch2009,Schucany2006,Dean1999}. Such pre-testing conditions the subsequent primary test (the Welch's t-test in our case) on the outcome of the pre-test, altering the true Type I error rate and distorting the statistical inference. Given the known robustness of the t-test (especially Welch's) to moderate deviations from normality, particularly with reasonable sample sizes ($J$) \parencite{Welch1947,Zimmerman2004}, and the established issues with pre-testing, we proceed directly with the Welch's t-test, relying on the CLT for justification.

\subsection{The TISCA Algorithm Workflow}
\label{subsec:tisca_workflow}

TISCA operates iteratively, executing batches of simulations and evaluating statistical power until a desired threshold is met. The researcher initiates the process by providing the following inputs:

\begin{itemize}
    \item \textbf{Simulation Function/Code:} A user-provided function or script that executes one full simulation run. This function must take the DGP settings as input and return a data frame containing the calculated performance metrics for all models being compared in that single simulation.
    \item \textbf{Performance Metrics:} A list of column names from the simulation output corresponding to the performance metrics to be formally compared (e.g., c("pehe\_modelA", "pehe\_modelB", "rmse\_ate\_modelA", "rmse\_ate\_modelB")). Comparisons are typically set up between the proposed model and one or more benchmarks for each metric.
    \item \textbf{Target Statistical Power} ($1-\beta$): The desired probability of detecting a true difference if it exists (the standard in the scientific community is 0.80).
    \item \textbf{Significance Level}($\alpha$): The threshold for statistical significance, controlling the Type I error rate (the standard in the scientific community is 0.05).
    \item \textbf{Minimum Detectable Effect Sizes} (MDEs, $\delta$): A vector specifying the smallest true difference in the mean of each performance metric (Proposed Model Mean - Benchmark Model Mean) that the researcher wants to be able to detect with the target power. This is a crucial parameter reflecting practical significance. For instance, $\delta_{PEHE} = -0.1$ might indicate the goal is to detect if the proposed model's average PEHE is at least 0.1 units lower than the benchmark's. These values can be informed by pilot runs (e.g., 50 simulations) or domain knowledge.
    \item \textbf{Comparison Pairs:} Specification of which model metrics to compare against each other (e.g., compare `pehe\_modelA` vs `pehe\_modelB`, `rmse\_ate\_modelA` vs `rmse\_ate\_modelB`).
    \item \textbf{Simulation Batch Size:} ($B$) The number of new simulations to run in each iteration before recalculating power (e.g., $B=50$).
    \item \textbf{Initial Simulation Count} ($J_0$): An optional starting number of simulations (e.g., $J_0=50$) to perform before the first power check. This ensures sufficient data for initial variance estimates. Defaults to `batch\_size`.
    \item \textbf{Multiple Testing Correction Method:} The chosen method to adjust p-values if multiple hypotheses are tested simultaneously (options: `"none"`, `"bonferroni"`, `"holm"`, `"BH"` (Benjamini-Hochberg)). Default is `"none"`.
\end{itemize}

TISCA operates iteratively, executing batches of simulations and evaluating statistical power until a desired threshold is met. The algorithm requires several inputs specifying the simulation setup, the desired statistical guarantees, and the comparisons of interest. It then proceeds according to the pseudo-code outlined in Algorithm~\ref{alg:tisca}.

\begin{algorithm}[htbp] 
 \caption{TISCA: Test-Informed Simulation Count Algorithm (Part 1 of 2)}
 \label{alg:tisca} 

 \SetKwInOut{KwIn}{Input}
 \SetKwInOut{KwOut}{Output}

 \KwIn{
    User simulation function $SimFunc(seed)$ returning performance metrics\;
    List of $K$ comparison pairs $Comparisons = \{ (metric_k, modelP_k, modelB_k) \}_{k=1}^K$\;
    Vector of $K$ Minimum Detectable Effect Sizes $MDEs = \{ \delta_k \}_{k=1}^K$\;
    Target statistical power $P_{target}$ (e.g., 0.80)\;
    Significance level $\alpha$ (e.g., 0.05)\;
    Simulation batch size $B$\;
    Initial simulation count $J_0$ (optional, e.g., 50)\;
    Multiple testing correction method $CorrMethod \in \{\text{"none"}, \text{"bonferroni"}, \text{"holm"}, \text{"BH"}\}$\;
 }

 \BlankLine

 \tcp{Initialization Phase}
 $J \leftarrow 0$\;
 $results\_agg \leftarrow \emptyset$ \tcp*{e.g., an empty dataframe}
 $P_{current} \leftarrow \mathbf{0}_{K}$ \tcp*{vector of $K$ zeros}

 \BlankLine

 \tcp{Optional Initial Run}
 \If{$J_0 > 0$ \textbf{and} $J_0 \ge B$}{
    \For{$i = 1$ \KwTo $J_0$}{
        $metrics\_run \leftarrow SimFunc(seed=i)$\;
        Append $metrics\_run$ to $results\_agg$\;
    }
    $J \leftarrow J_0$\;
    \tcp{Perform initial power calculation before main loop}
    $p_{values\_raw\_init} \leftarrow \mathbf{0}_{K}$\;
    $test\_stats\_init \leftarrow \mathbf{0}_{K}$\;
    \For{$k = 1$ \KwTo $K$}{
         \lIf{$sd_P > 0$ \textbf{and} $sd_B > 0$}{
              $P_{current}[k] \leftarrow \text{EstimateWelchPower}(J, J, sd_P, sd_B, \delta_k, \alpha)$
         }\lElse{
              $P_{current}[k] \leftarrow 0$
         }
    }
 }
\end{algorithm}

\begin{algorithm}[htbp] 
 \ContinuedFloat 
 \caption{TISCA: Test-Informed Simulation Count Algorithm (Part 2 of 2)} 

 \SetKwInOut{KwOut}{Output} 
 \KwOut{
    Final required simulation count $J_{final}$\;
    Vector of final raw p-values $P_{raw} = \{ p_{raw,k} \}_{k=1}^K$\;
    Vector of final adjusted p-values $P_{adj} = \{ p_{adj,k} \}_{k=1}^K$\;
    Vector of final test statistics $Stats = \{ stat_k \}_{k=1}^K$\;
    Vector of final estimated achieved powers $P_{achieved} = \{ p_{achieved,k} \}_{k=1}^K$\;
 }

 \BlankLine

 \tcp{Iterative Simulation and Power Check Loop}
 \While{$\min(P_{current}) < P_{target}$}{
    \tcp{Run a new batch of simulations}
    $start\_seed \leftarrow J + 1$\;
    $end\_seed \leftarrow J + B$\;
    \For{$i = start\_seed$ \KwTo $end\_seed$}{
        $metrics\_run \leftarrow SimFunc(seed=i)$\;
        Append $metrics\_run$ to $results\_agg$\;
    }
    $J \leftarrow J + B$\;

    \BlankLine
    \tcp{Perform tests and estimate power on current $J$ simulations}
    \For{$k = 1$ \KwTo $K$}{
         \lIf{$sd_P > 0$ \textbf{and} $sd_B > 0$}{ \tcp{Need valid SDs}
              $P_{current}[k] \leftarrow \text{EstimateWelchPower}(J, J, sd_P, sd_B, \delta_k, \alpha)$
         }\lElse{
              $P_{current}[k] \leftarrow 0$ \tcp{Cannot estimate power yet}
         }
    }
 } 

 \BlankLine
 \tcp{Output Phase}
 $J_{final} \leftarrow J$\;
 $P_{raw} \leftarrow p_{values\_raw}$ \tcp*{Final raw p-values}
 $P_{adj} \leftarrow p_{values\_adj}$ \tcp*{Final adjusted p-values}
 $Stats \leftarrow test\_stats$ \tcp*{Final test statistics}
 $P_{achieved} \leftarrow P_{current}$ \tcp*{Final estimated powers}

 \Return $J_{final}, P_{raw}, P_{adj}, Stats, P_{achieved}$\;

\end{algorithm}

This pseudo-code describes the iterative process: initializing, optionally running a pilot batch, then looping through batches of simulations, performing Welch's t-tests on the accumulated data, applying multiple testing corrections, estimating the current power for each comparison based on the observed variances and the target MDE, and stopping only when the power target is achieved for all comparisons. Helper functions like `SimFunc`, `Welch\_t\_test`, `AdjustPValues`, `StandardDeviation`, and `EstimateWelchPower` represent the underlying necessary computational steps detailed elsewhere in the text or implemented in the actual code.

This iterative process ensures that simulations continue precisely until the study has sufficient power to detect the pre-defined minimally important differences, providing a statistically justified stopping point and simulation count ($J$).

\subsection{Addressing Multiple Comparisons}
\label{subsec:multiple_comparisons}

Researchers often compare a new model against benchmarks across multiple performance metrics (e.g., PEHE, RMSE$_{ATE}$, Coverage) or against multiple different benchmarks simultaneously. Performing multiple hypothesis tests increases the probability of making at least one Type I error (a false positive finding) across the family of tests – the "multiple comparisons problem" \parencite{Streiner2011,Menon2019}.

TISCA acknowledges this issue by offering optional p-value adjustment procedures as an input parameter. The available methods include:
\begin{itemize}
    \item \textbf{Bonferroni correction:} A simple but often overly conservative method that controls the Family-Wise Error Rate (FWER) by multiplying each p-value by the number of tests (or equivalently, dividing $\alpha$ by the number of tests) \parencite{Dunn1961}.
    \item \textbf{Holm's method (Holm-Bonferroni):} A step-down procedure that also controls the FWER but is uniformly more powerful than the standard Bonferroni correction \parencite{Holm1979}.
    \item \textbf{Benjamini-Hochberg (BH) procedure:} Controls the False Discovery Rate (FDR) – the expected proportion of rejected null hypotheses that are actually true \parencite{Benjamini1995}. This is generally less conservative and more powerful than FWER-controlling methods, making it suitable when controlling the proportion of false positives among the significant findings is the primary goal.
\end{itemize}
The choice of correction method depends on the researcher's specific goals and tolerance for Type I versus Type II errors. While these methods provide established ways to handle multiple tests, they are not perfect solutions; FWER methods can be overly strict, potentially masking true effects, while FDR control allows for some false positives among the declared significant results \parencite{Menon2019}. The default setting of `"none"` assumes that only a few tests will be performed (say two or three), yet if more tests are performed, then we advise researchers choose one of the adjustment procedures considering their trade-offs.

\subsection{Implementation and Practical Use}
\label{subsec:implementation}

TISCA is designed to be implemented as a flexible function, envisioned primarily within the R statistical environment, leveraging existing packages for Welch's t-tests (`stats::t.test`) and p-value adjustments (`stats::p.adjust`).

To facilitate the adoption and use of TISCA, particularly for researchers who may be less familiar with power analysis or R programming, we have developed a companion web-based tool. Hosted on \url{https://tisca-llm-app.streamlit.app/}, this tool features an interactive interface powered by a fine-tuned LLM, namey Gemini 2.5 Flash. The LLM is fine-tuned on the TISCA methodology and its R implementation details, allowing researchers to describe their simulation study setup and receive guidance on structuring their simulation code, choosing appropriate parameters (like MDEs based on pilot data), and interpreting the TISCA output. This aims to lower the barrier to entry for adopting more rigorous simulation practices.

\subsection{Advantages and Limitations}
\label{subsec:advantages_limitations}

This approach offers statistical rigor by replacing arbitrary choices of $J$ with a statistically principled method grounded in power analysis and formal hypothesis testing (Welch's t-test). It provides a justified simulation count, offering a clear, data-driven rationale for the number of simulations, thereby enhancing reproducibility and credibility. The method promotes efficiency by avoiding unnecessary computational costs; simulations stop once sufficient power is achieved, saving resources and time compared to using an overly large, fixed $J$. It also prevents underpowered studies by ensuring simulations run until the desired power is reached. Furthermore, it fosters a focus on effect size, requiring researchers to explicitly define the MDE and encouraging the consideration of practical significance alongside statistical significance. Finally, it offers flexibility, accommodating various performance metrics, multiple comparisons (with appropriate corrections), and user-defined simulation code.

Despite its advantages, this method has a few limitations. The resulting $J$ is highly dependent on the MDE chosen; specifying an extremely small MDE can lead to computationally prohibitive simulation counts, making careful consideration and justification of the MDE crucial. While potentially more efficient than over-simulation, the iterative power calculation itself adds some computational overhead, especially if power estimation involves many simulations. The overall cost still heavily depends on the complexity of a single simulation run. It also relies on certain assumptions, specifically the appropriateness of Welch's t-test and the adequacy of the CLT approximation for the distributions of performance metrics. While robust, severe violations of these assumptions could affect results. Lastly, the sequential testing nature, which involves repeated checks on accumulating data, means the stopping rule itself is data-dependent. While the final inference uses the full $J$ dataset and standard tests/corrections, and the implications for error rates are generally considered minor in the context of achieving a target power for a frequentist test based on pre-specified MDE and alpha \parencite{Jennison2000}, it does represent a departure from fixed-sample designs.

Despite these limitations, TISCA offers a substantial improvement over current common practices by embedding statistical rigor and efficiency directly into the design and execution of Monte Carlo simulation studies for treatment effect estimation model evaluation.

\section{Real Life Example: Revisiting \citeauthor{McJames2024} (\citeyear{McJames2024})}\label{Sec4}

To illustrate the practical application and utility of TISCA, we revisit the simulation study presented in \citeauthor{McJames2024} (\citeyear{McJames2024}). This recent work introduced the Multivariate Bayesian Causal Forest (MVBCF) model for estimating treatment effects for multiple outcomes simultaneously. Their evaluation included extensive Monte Carlo simulations, namely three different DGPs and 1000 replications for each, comparing MVBCF against relevant benchmarks, including standard Bayesian Causal Forests (BCF) \parencite{Hahn2020} applied separately to each outcome (denoted here as 'wsBCF' or simply 'BCF' where context is clear), Bayesian Additive Regression Trees (BART) \parencite{Chipman2010}, and a multivariate version of BART (MVBART) \parencite{Um2022}. By applying TISCA while replicating their a part of simulation studies, we aim to demonstrate how our algorithm could have provided formal statistical evidence for model comparisons and determined a sufficient, potentially more efficient, number of simulation replications.

Our analysis specifically focuses on the outcomes reported for the first DGP (DGP1) in \citeauthor{McJames2024} (\citeyear{McJames2024}) with a training sample size of n=500 observations and test size of 1000 observations. This focused scope is chosen deliberately. DGP1 was constructed by \citeauthor{McJames2024} (\citeyear{McJames2024}) to represent conditions where the assumptions behind the MVBCF model are met, namely where the prognostic baseline ($\mu$) and the treatment effect modification ($\tau$) components for the two outcome variables, $Y_1$ and $Y_2$, share influential predictors and exhibit similar functional dependencies. Evaluating performance in this "ideal" scenario provides a critical test case for the primary claims of the model's superiority. Furthermore, the detailed numerical results for the DGPs of the paper are directly available in the main text of the original publication (their Table 2) only for n=500, motivating us to use n=500, albeit \citeauthor{McJames2024} (\citeyear{McJames2024}) have also explored training sample sizes of 100 and 1000 (in all cases the test size remains 1000 observations). While \citeauthor{McJames2024} (\citeyear{McJames2024}) presented two additional DGPs, our purpose here is illustrative—to showcase TISCA's utility—rather than performing an exhaustive replication. Thus, concentrating on DGP1 allows for a clear demonstration while maintaining conciseness.

In DGP1, \citeauthor{McJames2024} (\citeyear{McJames2024}) generated ten covariates ($X_1, \dots, X_{10}$) with a mix of distributions intended to mimic their real-world Trends in International Mathematics and Science Study (TIMSS) data (to which they applied their proposed model in their paper): $X_1, \dots, X_5 \sim U(0, 1)$; $X_6, \dots, X_8 \sim \text{Bernoulli}(0.5)$; and $X_9, X_{10}$ as ordinal variables with five equally likely levels $\{0, 1, 2, 3, 4\}$. The treatment assignment $Z_i$ followed $P(Z_i=1) = X_{4,i}$, making $X_4$ an observed confounder. The outcomes $Y_{1,i}$ and $Y_{2,i}$ were generated according to the following functional forms (adapted from their Table 1 \parencite{McJames2024}):

\begin{align}
Y_{1,i} &= \underbrace{\left(300 + 110 \sin (\pi X_{1,i} X_{2,i}) + 180(X_{3,i} - 0.5)^2 + 100X_{4,i} + 120X_{6,i} + 10X_{9,i}\right)}_{\mu_1(\mathbf{X}_i)} \nonumber \\
&\quad + \underbrace{\left(20X_{4,i} + 20X_{5,i}\right)}_{\tau_1(\mathbf{X}_i)} Z_i + \epsilon_{1,i} \label{eq:dgp1_y1} \\
Y_{2,i} &= \underbrace{\left(300 + 90 \sin (\pi X_{1,i} X_{2,i}) + 220(X_{3,i} - 0.5)^2 + 140X_{4,i} + 80X_{6,i} + 10X_{9,i}\right)}_{\mu_2(\mathbf{X}_i)} \nonumber \\
&\quad + \underbrace{\left(10X_{4,i} + 30X_{5,i}\right)}_{\tau_2(\mathbf{X}_i)} Z_i + \epsilon_{2,i} \label{eq:dgp1_y2}
\end{align}
where the error terms $(\epsilon_{1,i}, \epsilon_{2,i})$ were drawn from a multivariate normal distribution $MVN(\mathbf{0}, \Sigma)$ with $\Sigma = 50^2 \mathbf{I}$, calibrating the signal-to-noise ratio based on residual variance observed in their target TIMSS dataset.

The subsequent DGPs served essentially as sensitivity analyses. DGP2 assessed robustness to unobserved confounding by modifying DGP1: the effect of $X_4$ on $\mu_2$ was removed, and $X_4$ was excluded from the available covariates. This setup induced confounding only for $Y_1$, allowing the authors to investigate if performance degradation was isolated, which was in their presented results \parencite{McJames2024}. On the other hand, DGP3 examined model flexibility by altering the functional forms within DGP1: different interaction terms were used in $\mu_1$, and different linear combinations of covariates were used for $\tau_1$ and $\tau_2$. This tested whether MVBCF's structure could adapt when the two outcomes necessitated distinct tree representations for their prognostic and treatment effect components, which it could, though not perfectly \parencite{McJames2024}. Given that DGPs 2 and 3 probe robustness rather than baseline performance under favorable conditions, focusing our TISCA demonstration on DGP1 remains the most direct way to illustrate its utility in rigorously comparing models and determining simulation sufficiency for core performance claims.

Table~\ref{tab:mcjames_results} reproduces the key performance metrics for DGP1 with n=500 from Table 2 in \citeauthor{McJames2024} (\citeyear{McJames2024}), focusing on the comparison between MVBCF, BCF (wsBCF), and MVBART for PEHE and 95\% coverage of the true treatment effect ($\tau (\mathbf{X})$).

The original study reported the following key results (Mean ± SE across 1000 simulations) for DGP1, n=500:

\begin{table}[h]
 \centering
 \caption{Selected Simulation Results from \citeauthor{McJames2024} (Table 2, DGP1, n=500). Values are Mean ± SE.}
 \label{tab:mcjames_results}
 \begin{tabular}{lcccc}
 \hline
 Metric & Outcome & MVBCF & BCF & MVBART \\ \hline
 PEHE on $\tau$ & Y1 & 9.05 ± 0.16 & 9.63 ± 0.16 & 10.29 ± 0.19 \\
                & Y2 & 9.40 ± 0.16 & 9.96 ± 0.16 & 10.83 ± 0.19 \\ \hline
 $\tau$ 95\% Coverage & Y1 & 0.96 ± 0.00 & 0.97 ± 0.00 & 0.98 ± 0.00 \\
                      & Y2 & 0.95 ± 0.00 & 0.96 ± 0.00 & 0.98 ± 0.00 \\
 \hline
 \end{tabular}
\end{table}

Based on these results, the authors concluded: "Looking at the PEHE results from DGP1... multivariate BCF clearly outperforms the other three methods when tasked with accurately predicting heterogeneity in the treatment effect $\tau$... The PEHE, bias, and coverage results from DGP1 in Table 2 tell a very similar story... multivariate BCF shows minimal bias... and the 95\% coverage rate is close to ideal." \parencite{McJames2024}.

While the point estimates suggest MVBCF performs favorably regarding PEHE and coverage, the differences are not remarkable, especially in comparison to the BCF model given that the actual ATE for this DGP is around 20 units. Without formal statistical testing, it remains ambiguous whether these observed differences, based on 1000 simulations, reflect genuine superiority or could be attributed to Monte Carlo variability. This ambiguity motivates the application of TISCA to formally test these comparisons.

For the replication of the simulation study for DGP1 (n=500), we focused on the following six key comparisons motivated by the original paper's discussion and results:
\begin{enumerate}
    \item MVBCF vs. BCF: 95\% $\tau$ Coverage for Y1 (`mvbcf\_tau\_951` vs `wsbcf\_tau\_951`)
    \item MVBCF vs. BCF: 95\% $\tau$ Coverage for Y2 (`mvbcf\_tau\_952` vs `wsbcf\_tau\_952`)
    \item MVBCF vs. BCF: PEHE for Y1 (`mvbcf\_pehe1` vs `bcf\_pehe1`)
    \item MVBCF vs. BCF: PEHE for Y2 (`mvbcf\_pehe2` vs `bcf\_pehe2`)
    \item MVBCF vs. MVBART: 95\% $\tau$ Coverage for Y1 (`mvbcf\_tau\_951` vs `mvbart\_tau\_951`)
    \item MVBCF vs. MVBART: 95\% $\tau$ Coverage for Y2 (`mvbcf\_tau\_952` vs `mvbart\_tau\_952`)
\end{enumerate}

We configured TISCA with standard parameters: a significance level $\alpha = 0.05$ and a target statistical power $1-\beta = 0.80$. Based on the observed differences in the original study and representing plausible thresholds for practical significance, we set the Minimum Detectable Effect Sizes (MDEs, denoted $\delta$) as follows:
\begin{itemize}
    \item For PEHE comparisons (MVBCF vs. BCF): $\delta = 0.5$ (reflecting the approximate observed difference, testing if MVBCF PEHE is lower).
    \item For Coverage comparisons (MVBCF vs. BCF/MVBART): $\delta = 0.015$ (slightly larger than the observed 0.01 difference vs BCF, smaller than the average 0.025 difference vs MVBART, testing if MVBCF coverage is closer to 0.95, hence potentially lower than BCF/MVBART's coverage which exceeded 0.95).
\end{itemize}
We performed TISCA's iterative process using a batch size $B=100$.

\begin{figure}[H] 
\centering
\includegraphics[scale=0.55]{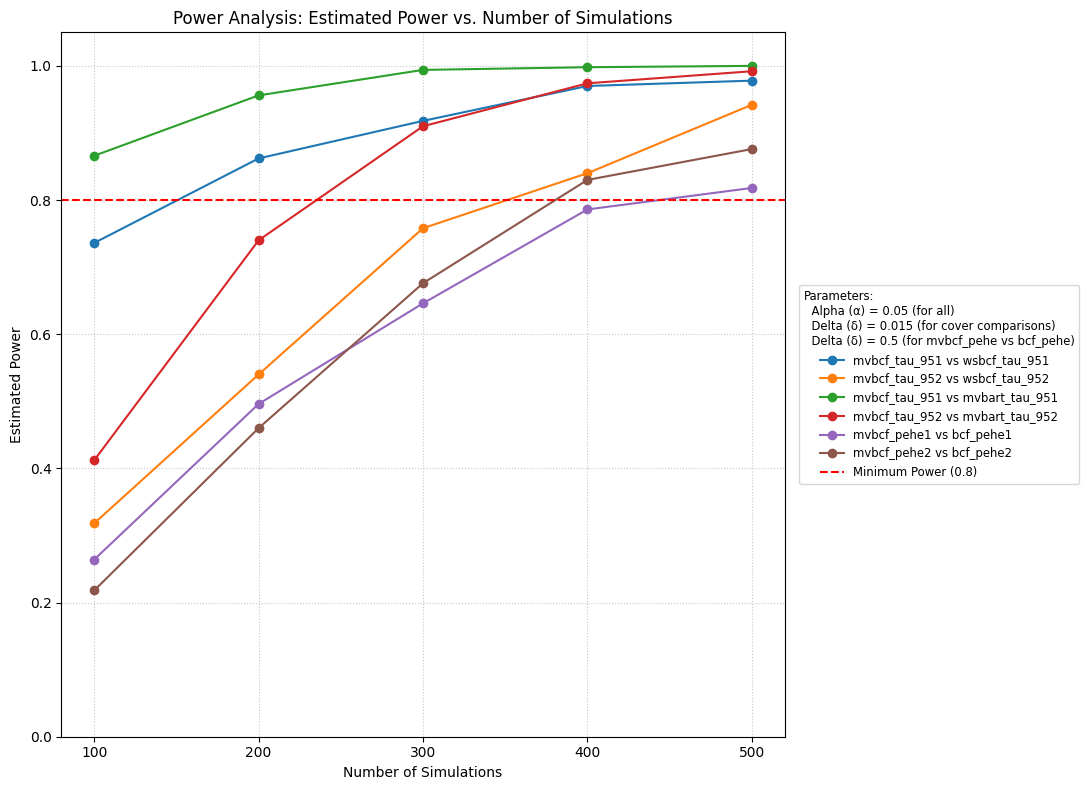} 
\caption{TISCA Power Analysis for \citeauthor{McJames2024} Simulation (DGP1, n=500). Estimated statistical power for detecting specified MDEs ($\delta$) for key comparisons as a function of the number of simulation replications ($J$). The red dashed line indicates the target power of 0.8.}
\label{Figure4}
\end{figure}

Figure~\ref{Figure4} shows the estimated statistical power for each of the six comparisons as TISCA accumulates simulation results. The power curves demonstrate how the ability to detect the specified MDE increases with the number of simulations $J$. Critically, for all six hypotheses, the estimated power reaches or exceeds the target threshold of 0.80 at or before $J=500$ simulations. This provides strong evidence that, for the chosen MDEs and significance level, performing 500 simulations would have been sufficient to achieve the desired statistical power for these specific comparisons.

\begin{figure}[H]
\centering
\includegraphics[scale=0.55]{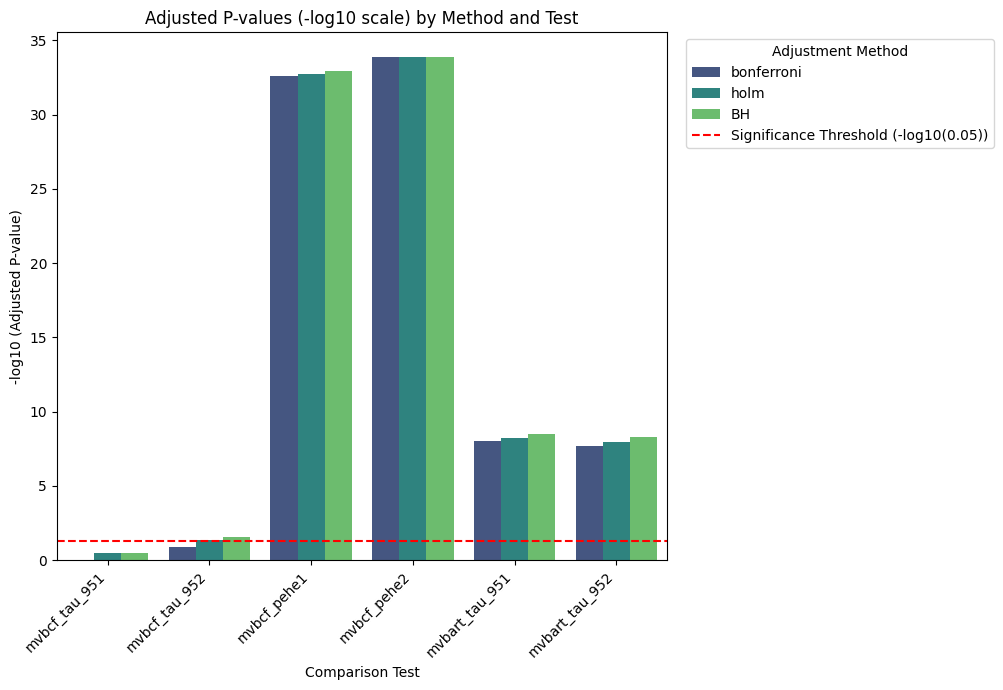} 
\caption{Original vs. Adjusted P-values from Welch's t-tests after $J=500$ simulations for \citeauthor{McJames2024} comparisons (DGP1, n=500). Values are shown on a -log10 scale (higher bars indicate smaller p-values/stronger evidence against the null hypothesis of no difference). The red dashed line represents the significance threshold ($\alpha = 0.05$, i.e., -log10(0.05) $\approx$ 1.3).}
\label{Figure5}
\end{figure}

Having determined that $J=500$ simulations suffice based on the power analysis, we examine the results of the Welch's t-tests performed by TISCA after these 500 runs. The raw p-values and p-values adjusted for multiple comparisons (using Bonferroni, Holm, and Benjamini-Hochberg methods for the family of 6 tests) are presented in Table~\ref{tab:pvalue_results}. Statistical significance is indicated using asterisks. Figure~\ref{Figure5} provides a visual representation of these results on a -log10 scale.

\begin{table}[H] 
\centering
\sisetup{ 
    table-format=1.2e-1, 
    table-space-text-post = $^{***}$, 
    table-number-alignment = center,
    retain-zero-exponent = false 
}
\small 
\begin{tabular}{l S S S S} 
\toprule
Comparison Test & {Raw p-value} & {Adj. (Bonf.)} & {Adj. (Holm)} & {Adj. (BH)} \\
\midrule
MVBCF vs BCF ($\tau$ Cov, Y1) & 3.40e-01 & 1.00e+00 & 3.40e-01 & 3.40e-01 \\
MVBCF vs BCF ($\tau$ Cov, Y2) & 2.28e-02$^{**}$ & 1.37e-01 & 4.55e-02$^{**}$ & 2.73e-02$^{**}$ \\
\midrule
MVBCF vs BCF (PEHE, Y1) & 3.88e-34$^{***}$ & 2.33e-33$^{***}$ & 1.94e-33$^{***}$ & 1.16e-33$^{***}$ \\
MVBCF vs BCF (PEHE, Y2) & 2.36e-35$^{***}$ & 1.42e-34$^{***}$ & 1.42e-34$^{***}$ & 1.42e-34$^{***}$ \\
\midrule
MVBCF vs MVBART ($\tau$ Cov, Y1) & 1.51e-09$^{***}$ & 9.06e-09$^{***}$ & 6.04e-09$^{***}$ & 3.02e-09$^{***}$ \\
MVBCF vs MVBART ($\tau$ Cov, Y2) & 3.42e-09$^{***}$ & 2.05e-08$^{***}$ & 1.03e-08$^{***}$ & 5.14e-09$^{***}$ \\
\bottomrule
\end{tabular}
\caption{Welch's t-test Results for Key Comparisons after $J=500$ Simulations (DGP1, n=500). Significance levels: ** $p < 0.05$, *** $p < 0.01$.}\label{tab:pvalue_results}
\end{table}

The \textit{p}-values, both unadjusted and adjusted, provide a nuanced picture. The extremely small p-values, both raw and adjusted, for the PEHE comparisons (mvbcf\_pehe1 vs bcf\_pehe1 and mvbcf\_pehe2 vs bcf\_pehe2) offer overwhelming statistical evidence that MVBCF achieves significantly lower PEHE than the standard BCF implementation in this scenario. This strongly supports the authors' claims regarding MVBCF’s accuracy in estimating heterogeneous effects. Comparisons of $\tau$ coverage between MVBCF and MVBART (mvbart\_tau\_951 and mvbart\_tau\_952) also yield highly significant p-values across all adjustment methods, indicating a statistically significant difference in coverage properties between these two multivariate approaches in this setting. However, the results for $\tau$ coverage comparison between MVBCF and BCF are mixed. For Y1 (mvbcf\_tau\_951), the difference is not statistically significant ($p > 0.05$) regardless of adjustment. For Y2 (mvbcf\_tau\_952), the raw p-value (0.023) is below 0.05 and remains significant after Holm and BH correction, but not after the stricter Bonferroni correction, suggesting weaker evidence for a difference in coverage between MVBCF and BCF compared to the other tested differences. Overall, TISCA’s analysis largely substantiates the authors’ claims regarding MVBCF’s superior PEHE performance in DGP1, but adds nuance by highlighting that the evidence for superior or different coverage performance compared to standard BCF is less conclusive, particularly for outcome Y1.

Nonetheless, perhaps the most salient finding from this retrospective TISCA application is the determination that $J=500$ simulations were sufficient to achieve 80\% power for detecting the specified effect sizes across all six key comparisons. The original study performed $J=1000$ simulations. While performing additional simulations generally increases power and precision, TISCA demonstrates that, for the stated goals (detecting MDEs of 0.5 for PEHE and 0.015 for coverage with 80\% power), the final 500 simulations performed by \citeauthor{McJames2024} (\citeyear{McJames2024}) were potentially redundant.

This redundancy has significant practical implications. For our simulation studies replication, it took on average 10 hours to run 100 simulations using an Intel(R) Xeon(R) CPU @ 2.20GHz on Google Colab; thus, running their full set of simulations would require approximately 100 hours of computational time. Halving the number of simulations for DGP1 (and potentially for other DGPs, had TISCA been applied there) could have saved roughly \textbf{50 hours of computation}. This translates directly to increased research efficiency, as the saved computational resources and researcher time could be redirected towards other valuable activities, such as exploring additional DGPs, testing sensitivity to hyperparameters, analyzing more real-world data, or developing further methodological improvements. Additionally, it contributes to enhanced sustainability, as reducing unnecessary computation lowers the energy consumption and associated environmental footprint of research activities, supporting more sustainable scientific practices—a growing concern in computationally intensive fields \parencite{Lannelongue2021}. Finally, this approach leads to improved reliability and comparability; by providing a statistically justified simulation count, TISCA enhances the reliability of the study’s conclusions regarding statistical power. Furthermore, widespread adoption of such principled approaches, rather than arbitrary choices of $J$, would significantly improve the comparability of simulation results across different studies in the treatment effect estimation literature.

In summary, this case study demonstrates how TISCA can be applied to real-world simulation studies to provide formal statistical comparisons and determine an adequate number of simulations. It confirms key findings of the original study (\citeauthor{McJames2024}) (\citeyear{McJames2024}) regarding PEHE while adding statistical nuance to coverage comparisons, and critically highlights the potential for significant gains in computational efficiency and research sustainability without sacrificing statistical rigor.

\section{Conclusion}\label{Sec5}

This paper addressed critical methodological shortcomings prevalent in the evaluation of treatment effect estimators via Monte Carlo simulation studies in the current literature: the widespread absence of formal statistical hypothesis testing for model comparisons and the often arbitrary selection of the number of simulation replications ($J$). As evidenced by our bibliometric analysis, current practices exhibit significant heterogeneity in the choice of $J$, frequently relying on convention (such as $J=1000$) rather than statistical principles. This lack of rigor can lead to underpowered studies, unsubstantiated claims of model superiority, inefficient use of computational resources, and difficulties in comparing findings across the literature---concerns that are amplified in the current era of rapidly proliferating AI-driven research where robust validation is paramount \parencite{Goldkuhle2024,doVale2024}.

As a solution, we introduced the Test-Informed Simulation Count Algorithm (TISCA). TISCA provides a systematic and statistically grounded framework designed to integrate rigorous hypothesis testing directly into the simulation workflow. By leveraging the robust Welch's t-test and principles of statistical power analysis, TISCA iteratively performs simulations until a user-specified power level (e.g., 0.80) is achieved for detecting pre-defined Minimum Detectable Effect Sizes (MDEs) at a given significance level ($\alpha$, e.g., 0.05). It explicitly accounts for potential unequal variances between comparison groups and incorporates options for multiple testing corrections (Bonferroni Correction \parencite{Dunn1961}, Holm's method \parencite{Holm1979}, and Benjamini-Hochberg procedure \parencite{Benjamini1995}) when necessary.

The practical utility and benefits of TISCA were demonstrated through a case study revisiting the simulation experiments of \citeauthor{McJames2024} (\citeyear{McJames2024}). Our retrospective application revealed that for their primary simulation scenario, achieving 80\% power for detecting meaningful differences in PEHE and coverage required only half the number of simulations ($J=500$) originally performed ($J=1000$). This finding highlights the potential for substantial gains in computational efficiency---saving valuable research time and resources---and contributes to more environmentally sustainable research practices \parencite{Lannelongue2021}. Furthermore, the TISCA analysis provided formal statistical evidence that largely supported the original study's conclusions regarding the superior PEHE of their proposed MVBCF model, while adding important nuance regarding the statistical significance of observed differences in coverage metrics compared to benchmarks.

By adopting TISCA, researchers can move beyond arbitrary choices for $J$, ensuring their simulation studies are adequately powered to detect effects deemed practically significant, while simultaneously avoiding wasteful over-simulation. This methodology fosters greater transparency, reproducibility, and comparability within the field of treatment effect estimation. The provision of a statistically justified simulation count strengthens the credibility of research findings and facilitates more reliable assessments of novel estimators. To aid adoption, we have also introduced plans for a companion web-based tool leveraging a fine-tuned LLM to guide researchers in applying TISCA to their specific simulation setups.

While TISCA offers significant advantages, we acknowledge its limitations, primarily the crucial dependence on the thoughtful specification of the MDE and the computational overhead associated with iterative power estimation. Future research could explore extensions of TISCA, such as incorporating alternative robust statistical tests suitable for different types of performance metrics or comparison scenarios (e.g., non-inferiority testing, multiple comparisons with a control). Further investigation into optimizing the power estimation step through the use of C++ code instead of R code and refining the guidance provided by the LLM tool also represent valuable avenues for development.

In conclusion, TISCA provides a needed methodological advancement for conducting simulation studies in causal inference and beyond. By embedding statistical rigor and efficiency into the core of the evaluation process, it offers a pathway towards more reliable, resource-conscious, and comparable research, ultimately strengthening the foundation upon which new methods for treatment effect estimation are developed, validated, and compared.

\section*{Supplementary Research Material and Code}
The code for TISCA as well as the code used Section \ref{Sec4} can be found in the following GitHub repository: \url{https://github.com/hugogobato/Test-Informed-Simulation-Count-Algorithm-TISCA}. Additionally, the website to facilitate the use of TISCA by researchers can be found here: \url{https://tisca-llm-app.streamlit.app/}.

\printbibliography

\end{document}